\documentclass[]{spie}  

\usepackage{amsmath,amsfonts,amssymb}
\usepackage{graphicx}
\usepackage[colorlinks=true, allcolors=blue]{hyperref}

\usepackage{xcolor}
 
\newcommand{\citefig}[1]{(taken from \cite{#1}, with permission)}  
\newcommand{\citefigadapt}[1]{(adapted from \cite{#1}, with permission)}  
\definecolor{color1}{rgb}{0.565, 0.0, 0.125}
\definecolor{color2}{rgb}{0.169, 0.588, 0.714}

\title{Gamma-Ray Burst Polarimetry with the POLAR and POLAR-2 missions}

\author[a]{Nicolas~De~Angelis}  
\author[b]{Philipp~Azzarello}  
\author[b]{Franck~Cadoux}
\author[c]{Kurt~Dittrich}
\author[b]{Yannick~Favre}
\author[c]{Jochen~Greiner}  
\author[b]{Johannes~Hulsman}  
\author[b]{Coralie~Husi}
\author[d]{Merlin~Kole}  
\author[e]{Hancheng~Li}  
\author[f]{Slawomir~Mianowski}  
\author[b]{Gabriel~Pelleriti}
\author[f]{Agnieszka~Pollo}  
\author[e]{Nicolas~Produit}  
\author[f]{Dominik~Rybka}  
\author[g]{Jianchao~Sun}  
\author[b]{Xin~Wu}  
\author[g,h]{Shuang-Nan~Zhang}  

\affil[a]{INAF-IAPS, via del Fosso del Cavaliere 100, 00133 Rome, Italy}
\affil[b]{DPNC, University of Geneva, Quai Ernest-Ansermet 24, 1205 Geneva, Switzerland}
\affil[c]{Max-Planck Institute for Extraterrestrial Physics, Giessenbachstr. 1, 85748 Garching, Germany}
\affil[d]{University of New Hampshire, Space Science Center, University of New Hampshire, Durham, NH 03824, USA}
\affil[e]{Geneva Observatory, ISDC, University of Geneva, Chemin d’Ecogia 16, 1290 Versoix, Switzerland}
\affil[f]{National Centre for Nuclear Research, ul. A. Soltana 7, 05-400 Otwock, Swierk, Poland}
\affil[g]{Key Laboratory of Particle Astrophysics, Institute of High Energy Physics, Chinese Academy of Sciences, 100049 Beijing, China}
\affil[h]{University of Chinese Academy of Sciences, 100049 Beijing, China}

\authorinfo{Further author information: (Send correspondence to N.D.A.)\\N.D.A.: E-mail: \href{mailto:nicolas.deangelis@inaf.it}{nicolas.deangelis@inaf.it}}

\begin{document} 
\maketitle

\begin{abstract}
Gamma-Ray Bursts (GRBs) are among the most powerful and violent events in the Universe. Despite over half a century of observations of these transient sources, many open questions remain about their nature and the physical emission mechanisms at play. Polarization measurements of the GRB prompt $\gamma$-ray emission have long been theorized to be able to answer most of these questions. Early polarization measurements did not allow to draw clear conclusions because of limited significance.

With the aim of better characterizing the polarization of these prompt emissions, a compact Gamma-Ray polarimeter called POLAR has been sent to space as part of the Tiangong-2 Chinese space lab for 6 months of operations starting September 2016. The instrument detected 55 GRBs as well as several pulsars. Time integrated polarization analysis of the 14 brightest detected GRBs has shown that the prompt emission is lowly polarized or fully unpolarized. However, time-resolved analysis depicted strong hints of an evolving polarization angle within single pulses, washing out the polarization degree in time-integrated analyses. Energy-resolved polarization analysis has shown no constraining results due to limited statistics. Hence, a more sensitive $\gamma$-ray polarimeter is required to perform detailed energy and time-resolved polarization analysis of the prompt $\gamma$-ray emission of GRBs.
        
Based on the success of the POLAR mission, a larger-scale instrument, approved for launch to the China Space Station (CSS) in 2027, is currently being developed by a Swiss, Chinese, Polish, and German collaboration. Thanks to its large sensitivity in the 20-800~keV range, POLAR-2 will produce polarization measurements of at least 50 GRBs per year with a precision equal to or higher than the best results published by POLAR, allowing for good quality time and energy resolved analysis. Furthermore, thanks to its large effective area which exceeds $2000\,\mathrm{cm^2}$ at 100~keV, POLAR-2 will be able to observe faint GRBs such as 170817A and will be capable of sending alerts of such transients, including localization information to the ground within seconds to minutes. POLAR-2 thereby not only aims to make the prompt polarization a standard observable, but it will additionally play an important role in multi-messenger observations.

The scientific results of the POLAR mission will be presented, followed by a discussion about the POLAR-2 mission, the future of GRB's prompt emission polarimetry.

\end{abstract}


\section{INTRODUCTION: GRB PROMPT EMISSION POLARIMETRY}\label{sec:intro}

Gamma-Ray Bursts (GRBs) are among the most powerful events in the Universe. They were discovered in 1967 by the Vela satellites \cite{Klebesadel73, Cline73}, whose main purpose was to monitor human nuclear activities in space to ensure the application of the Test Ban Treaty signed in 1963. GRBs consist of a bright prompt emission in the $\gamma$-ray band, followed by an afterglow spanning all electromagnetic wavelengths discovered later on \cite{Costa1997_xray_afterglow, Paradijs1997_optical_afterglow}. These events are uniformly distributed in the sky and are of extragalactic origin. They can be sub-classified into two categories based on the duration of their prompt emission $T_{90}$, with a conventional separation at 2~seconds. Short bursts can originate from NS-NS or NS-BH mergers, while long ones are thought to originate from the death of supermassive stars. The former can also be detected by Gravitational Waves observatories, as shown by the multi-messenger detection of GW170817 and GRB170817A \cite{GW170817_ligovirgo, Abbott_2017_counterpart}.

Despite almost 60 years of temporal, spectral, and localization studies of GRBs, they remain poorly understood. Current open questions about these powerful sources regard the emission mechanisms responsible for these bright cosmic events, the geometry of the ultra-relativistic ejecta causing the prompt emission through internal shocks and the afterglow through external shocks, the configuration of the magnetic fields, as well as the way the energy is being dissipated. Polarimetry of the prompt emission of GRBs is thought to be able to bring substantial knowledge in all of these key topics and therefore to push forward our understanding of these transient sources \cite{GRBpol_review_merlin_ramandeep_joni, Ramandeep_linpol_GRB}. Measuring the polarization of the bursts is indeed a very powerful tool to probe the emission mechanism at play as well as the magnetic field and jet geometry.

Several polarimetric analyses of GRBs were attempted in the early 2000's, providing very contradictory results, sometimes even re-analyzing the same burst with data from the same mission \cite{MCCONNELL20171}. None of these by-product polarimeters provided reliable results as they were neither designed or calibrated for performing polarimetric measurements. The first instrument designed with the aim of measuring the polarization of GRBs in the $\gamma$ band was GAP \cite{GAP}, which only reported polarization results for 7 bursts with limited polarimetric significance. With the aim of providing a catalog of GRB prompt emission polarization measurements with higher significance, the POLAR detector was developed and launched in 2016, followed by its successor, POLAR-2, currently being developed for a launch in 2027. We summarize in this manuscript the results achieved by the POLAR observations as well as the development and status of the POLAR-2 mission.

\section{OUTCOMES FROM THE POLAR MISSION}\label{sec:polar}
\subsection{POLAR Instrument Design and Operation}\label{subsec:polar-instr}

\begin{figure}[h!]
\centering
 \includegraphics[height=.35\textwidth]{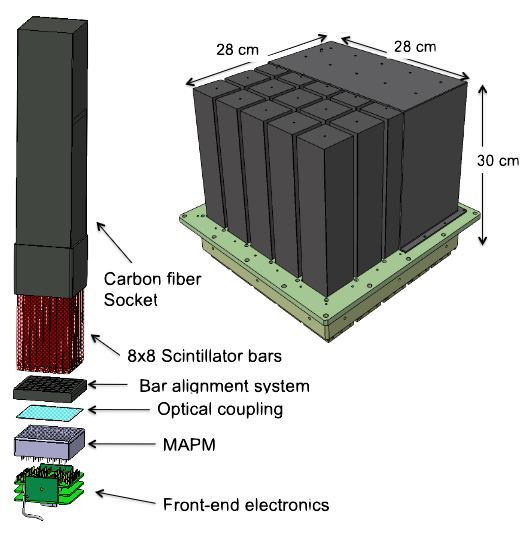}\hspace*{0.2cm}\includegraphics[height=.35\textwidth]{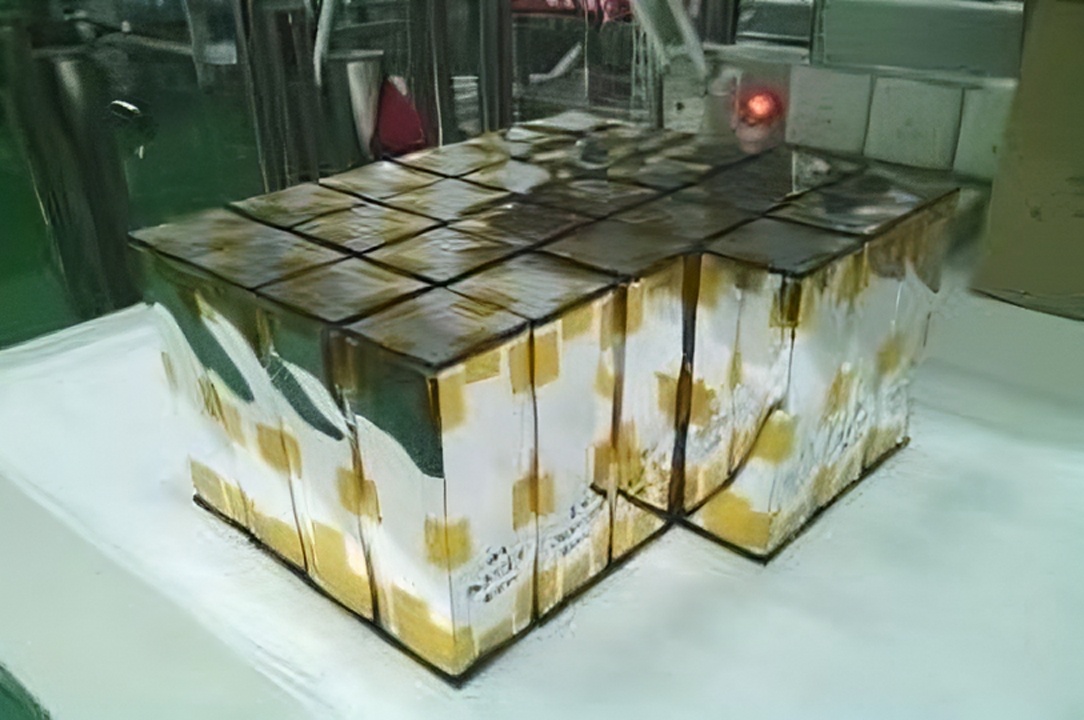}
 \caption{\textbf{Left:} POLAR's CAD design and exploded view of a polarimeter module \citefig{Estella_thesis}. \textbf{Right:} POLAR scintillator targets wrapped with 3M Vikuti Enhanced Specular Reflector films.}
 \label{fig:POLAR}
\end{figure}

In order to provide the first trustworthy catalog of GRB prompt polarimetry measurements, a dedicated polarimeter was designed by the POLAR collaboration. As a single-phase Compton polarimeter, POLAR made use of the modulated dependency between the azimuthal Compton scattering direction of a photon and its polarization vector, ruled by the Klein-Nishina cross section \cite{KN_cross_section_paper}, to infer the polarization parameters\footnote{Namely the Polarization Degree (PD) and Angle (PA).} of a given source. It consisted of a 40$\times$40 array of elongated scintillator bars divided into 25 modules each read out by a 64-channel Multi-Anode PhotoMultiplier Tube (MAPMT) from Hamamatsu. The scintillator material was chosen to be Eljen's PVT-based EJ-248M\footnote{Eljen EJ-248M Product Page -- \url{https://eljentechnology.com/products/plastic-scintillators/ej-244-ej-248-ej-244m-ej-248m}, consulted on 16 July 2025} due to its low-Z nature, well optimized for Compton scattering to be dominant down to lower energies (tens of keV) compared to a higher-Z inorganic scintillating materials. An exploded CAD design of a POLAR module as well as a picture of assembled polarimeter modules are shown in Figure \ref{fig:POLAR}, while \cite{POLARinstr} provides a detailed description of the POLAR instrument design and construction.

The POLAR instrument was launched together with the Tiangong-2 Chinese space lab on the 15\textsuperscript{th} of September 2016, onboard a Long March 2F rocket from the Jiuquan Satellite Launch Center. It has been operated for about 6 months and detected 55 GRBs as well as other sources in the 50-500~keV band. It had a half-sky Field of View with a polarimetric effective area of about $300~\textrm{cm}^2$ at 400~keV.

\subsection{The POLAR Results}\label{subsec:polar-res}

\begin{figure}[h!]
\centering
 \includegraphics[height=.4\textwidth]{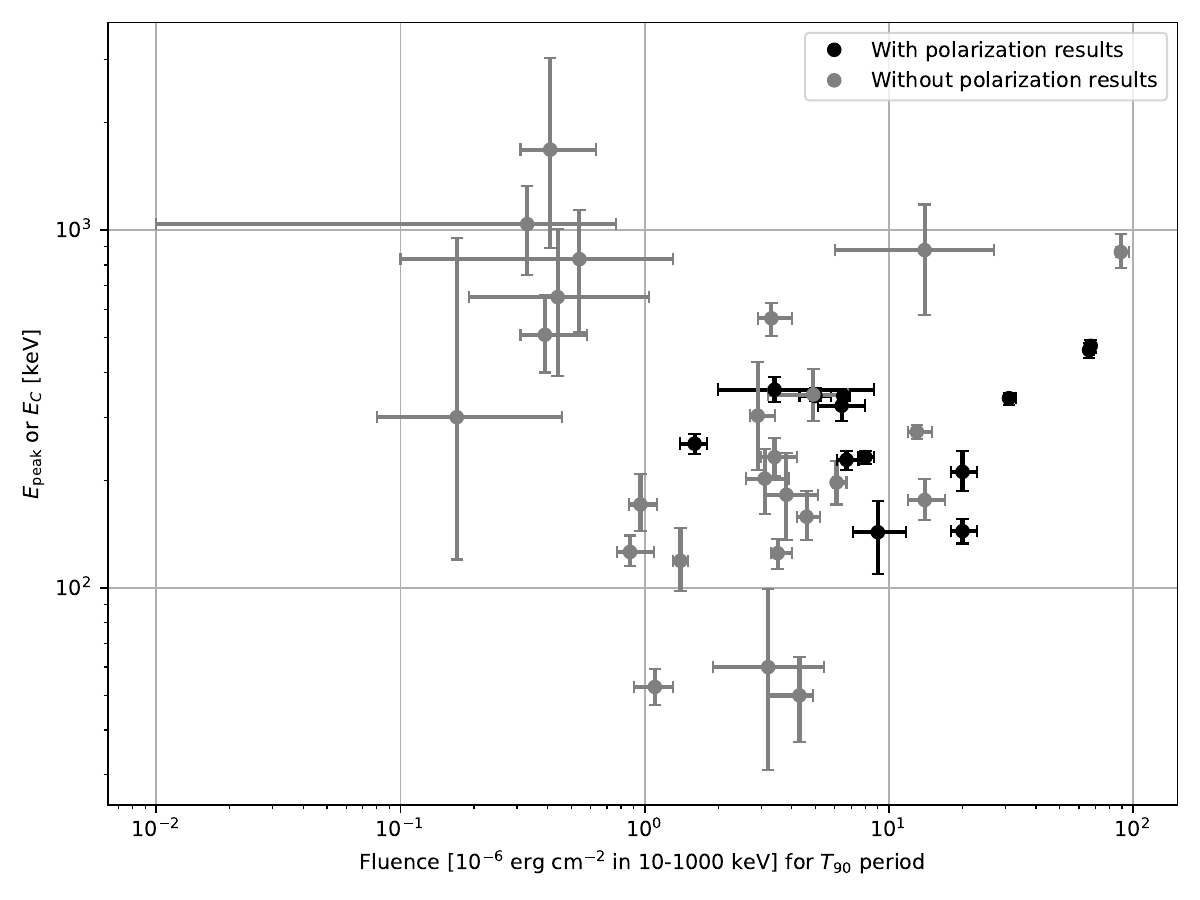}\hspace*{0.2cm}\includegraphics[height=.4\textwidth]{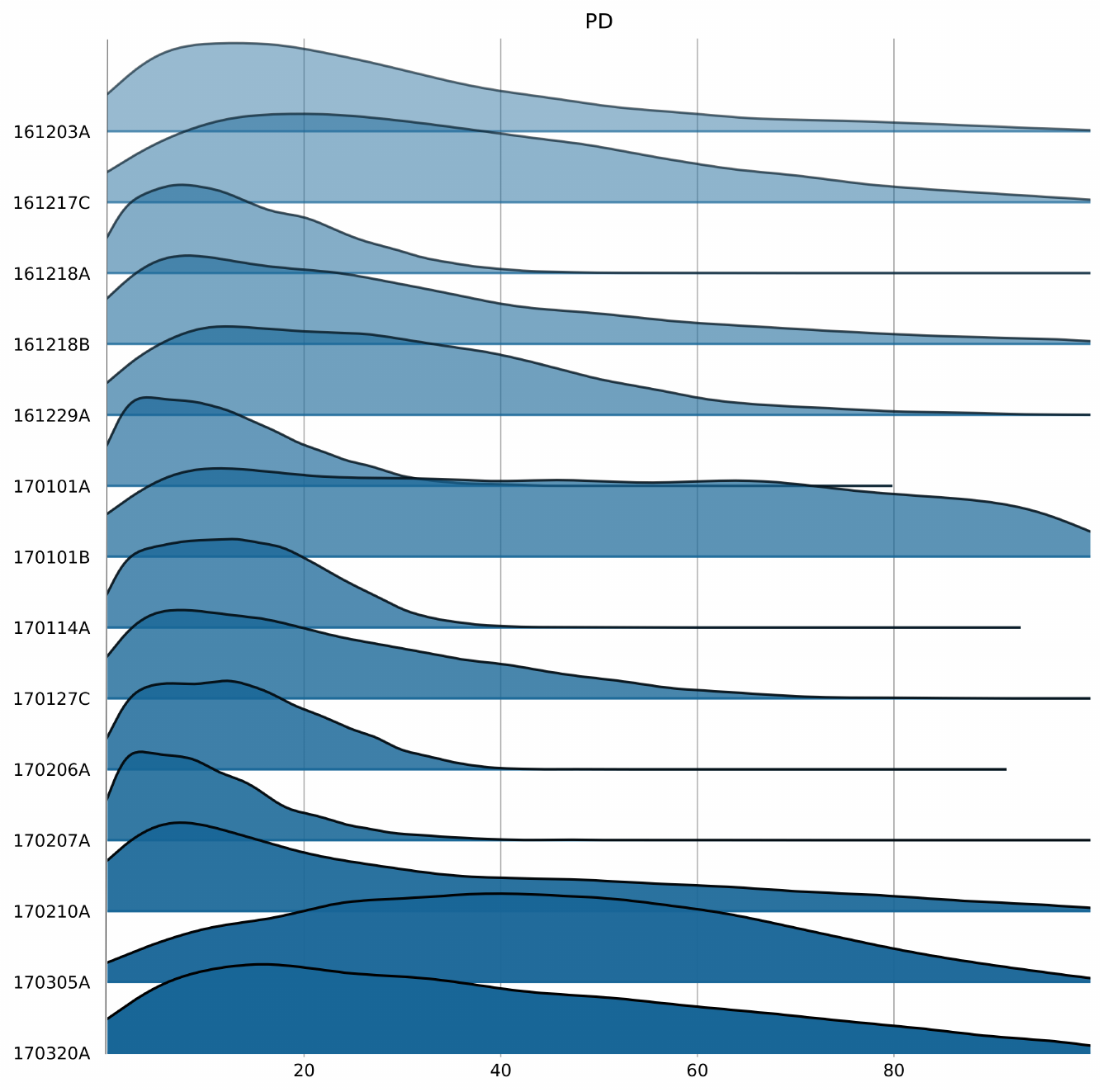}
 \caption{\textbf{Left:} E\textsubscript{peak} or E\textsubscript{c} [keV] (\textit{depending on spectral model: Band function \cite{Band93} or cutoff powerlaw}) as a function of the burst fluence in the 10-1000~keV band and $T_{90}$ time interval for the 38 GRBs detected by POLAR with an off-axis angle $<90^\circ$. The markers' color indicates whether the burst was analyzed for polarization. \textbf{Right:} Posterior distributions of the PD [\%] for the 14~GRBs for which polarization analysis has been performed in the POLAR catalog paper \citefig{POLAR_catalog}.}
 \label{fig:POLAR_results}
\end{figure}

A first analysis of the five brightest GRBs detected by POLAR was first performed \cite{POLAR_nature_astronomy}, followed by a refined analysis leading to the publication of a polarimetric catalog of 14 bursts \cite{POLAR_catalog}. The spectro-polarimetric analysis of these bursts was performed using the threeML analysis framework \cite{threeML}, which allowed to perform multi-instrument joined analysis for the spectral part using Fermi-GBM and Swift-BAT data for jointly observed events. The analysis method, described in \cite{POLAR_catalog}, consisted of parallelized forward-folding of the spectral and polarization parameters through the instrument's spectral and polarimetric responses, respectively, and comparing the results to the observation in order to find the best values of the spectro-polarimetric parameters of the model. The resulting posterior distributions of the polarization degree for the 14 analyzed bursts showed a low level of polarization for most of the bursts, compatible with a lowly polarized or even completely unpolarized flux, as can be appreciated in Figure \ref{fig:POLAR_results}.

The low level of polarization observed for these bursts could, however, hide a more complex behavior. This was hinted by a time-resolved analysis performed on the brightest burst of the catalog, GRB170114A, which showed a possible rotating polarization angle with a moderate polarization degree that would end up washed out by the integration in non-resolved analyses \cite{time_resolved_POLAR}. An energy-resolved analysis was also performed on this catalog \cite{ICRC23_POLAR_NDA, Energy_resolved_paperNDA}, but the limited statistics available in the POLAR observation did not allow us to conclude on any significant energy dependence of the polarization parameters. A more sensitive dedicated GRB polarimeter is therefore required for a deep understanding of the polarimetric behavior of the prompt emission of GRBs. A new generation Compton polarimeter, POLAR-2, is therefore under construction as described in Section \ref{sec:polar-2}.

In addition to its contribution to GRB science, it should be noted that POLAR was able to observe other classes of sources. First of all, both spectral and polarimetric phase-resolved analysis of the Crab pulsar was performed, showing results consistent with most of the past missions that already conducted such an observation in similar energy bands \cite{Crab_spec_POLAR, Crab_pol_ICRC_POLAR, Crab_pol_POLAR}. Moreover, POLAR detected 5 B-class and 11 C-class solar flares during its operation \cite{PING202087}. These flares were jointly detected by GOES and RHESSI, but were too faint to allow their polarimetric analysis. Finally, it is worth mentioning that POLAR also detected a magnetar-like transient event spectrally modeled with "\texttt{bbody+powerlaw}" \cite{koziol2021polarGCN}.

\newpage
\section{DESIGN AND DEVELOPMENT OF POLAR-2}\label{sec:polar-2}

Despite bringing exciting results, POLAR showed the need for the development of a more sensitive GRB-dedicated polarimeter to overcome the statistical limitations and perform detailed energy and time resolved polarimetric studies of the prompt emission from GRBs. Based on the POLAR legacy, an upgraded Compton polarimeter called POLAR-2 is under development with a foreseen launch in 2027. We summarize here the instrument's design and calibration as well as the space qualification campaign that were already conducted at module and component level.

\subsection{Instrument and Module Design}\label{subsec:instr_module_design}

\begin{figure}[h!]
\centering
 \includegraphics[height=.3\textwidth]{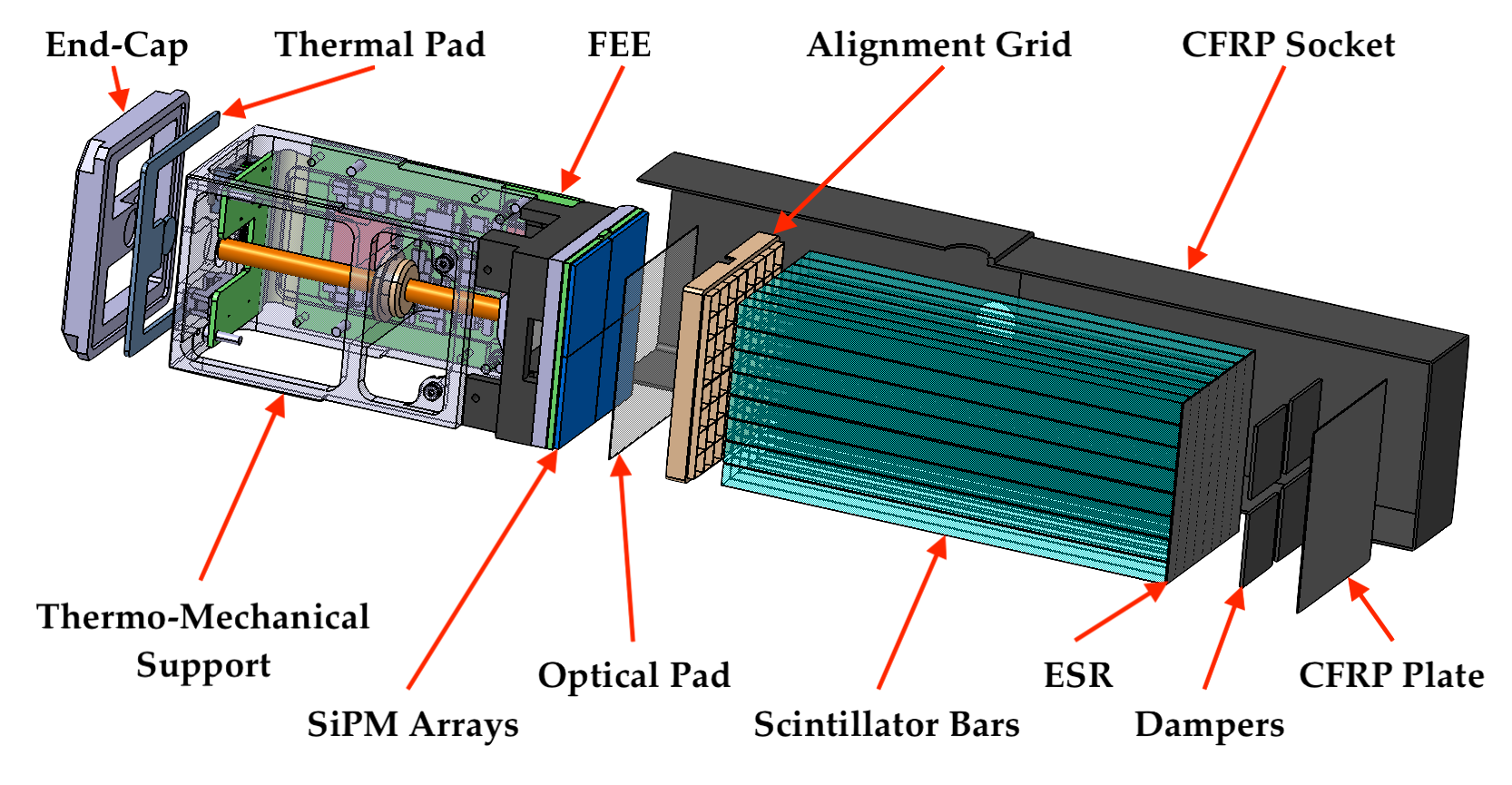}\hspace*{0.2cm}\includegraphics[height=.3\textwidth]{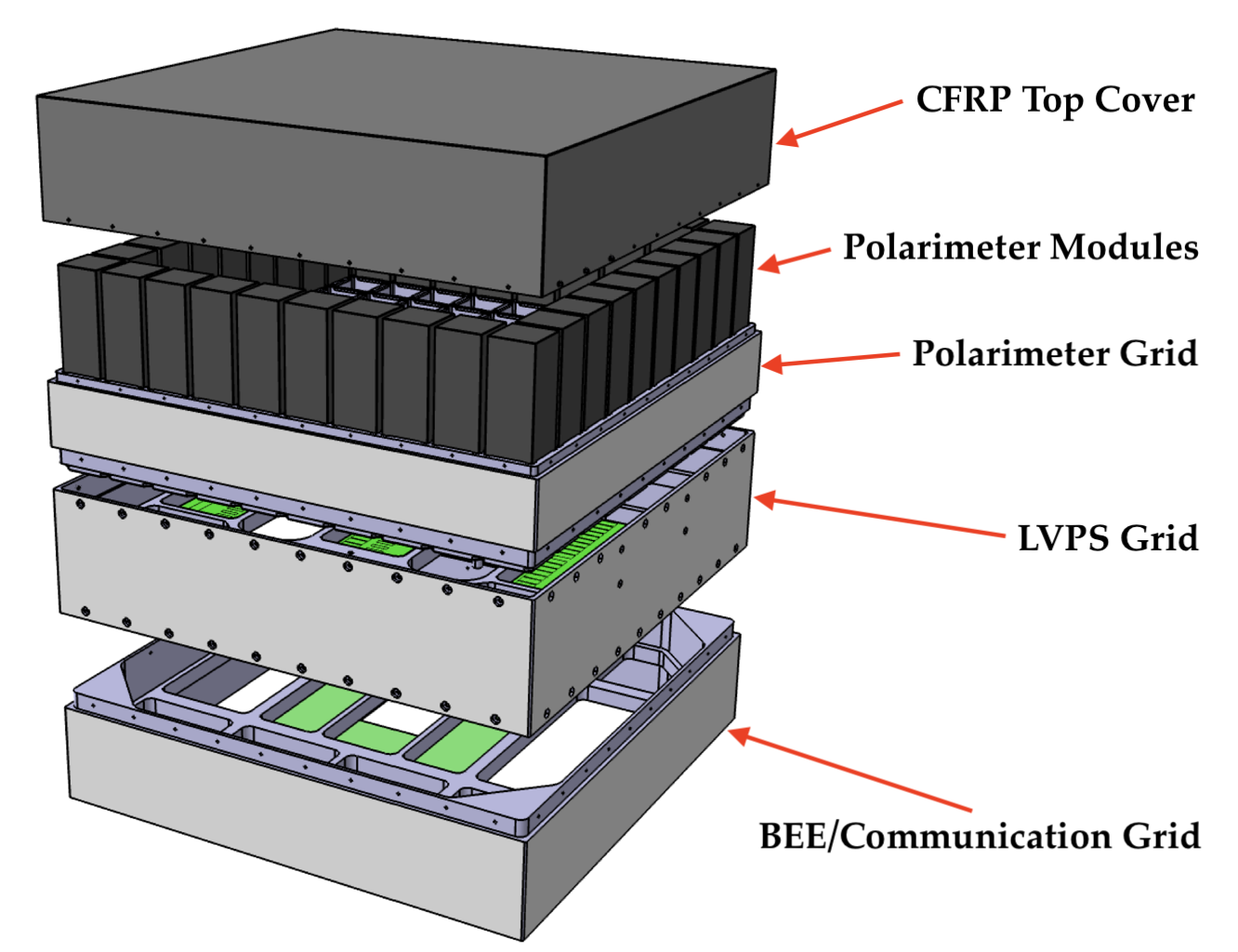}
 \caption{\textbf{Left:} Exploded CAD model of a POLAR-2 polarimeter module \citefig{NDA_thesis}. \textbf{Right:} Exploded CAD model of the POLAR-2 instrument \citefig{NDA_thesis}.}
 \label{fig:POLAR-2_CAD}
\end{figure}

POLAR-2 carries forward the modular philosophy of POLAR, expanding the array from 25 to 100 detector modules. The module itself is still based on an 8$\times$8 array of wrapped plastic scintillators read out by light sensors, although several upgrades have been brought to the design to improve its sensitivity. First of all, the scintillator bars have been shorten from 176~mm down to 125~mm in order to improve the signal-to-noise ratio, as the background is proportional to the scintillator volume while the source flux is proportional to the surface of the scintillators in the section orthogonal to the burst direction \cite{SNR_TN}. In addition to this, the readout sensor has been upgraded from an MAPMT to an array of Silicon PhotoMultipliers (SiPMs) with a photo-detection efficiency twice higher in the scintillator emission range (400-500~nm). This combined with an optimization of the opto-mechanical design of the module to reduce dead spaces and improve the efficiency of the wrapping led to an increase of the module light yield (amount of optical light collected by the sensors per unit of deposited energy) by over a factor 5 \cite{OptSim}. This design optimization was carried through a thorough optical characterization of all the elements comprising the module combined with a detailed optical simulation of the polarimeter module to fully understand its behavior\cite{OptSim}. Exploded computer-aided design views of both the polarimeter module and the entire instrument are shown in Figure \ref{fig:POLAR-2_CAD}.

A dedicated front-end electronics based on the CITIROC-1A ASIC was designed for POLAR-2\cite{FEEpaper}. It allows to read out 64 scintillator bars and apply a coincidence logic for reading coincident events from Compton scattered photons. This electronics can also be used to read out other crystals, and will be used coupled to GAGG segmented scintillators in POLAR-2's Broad-band Spectrometer (BSD)\cite{BSD} as well as in eXTP's Wide-field and Wide-band Camera (W2C)\cite{new_eXTP_instr}. The plastic material was chosen to be the same as for POLAR, i.e. EJ-248M from Eljen, since other plastics such as EJ-200 shown a lower light yield despite a higher scintillation efficiency due to more interface losses because of a rougher surface\cite{OptSim}. More details on the instrument and module design can be found in\cite{ICRC21_POLAR-2_NDA, ICRC21_POLAR-2_MK, ICRC23_POLAR-2_Produit, NDA_thesis}.

\subsection{Polarimeter Module Calibration}\label{subsec:calib}

A calibration setup have been designed in the POLAR-2 laboratory at CERN based on an X-Y motorized scanning system to which is attached a custom lead block hosting a $^{241}$Am source. The system allows to irradiate a given channel either with the direct collimated flux from the source, or with a 90$^\circ$ scattered polarized flux. The X-Y system is used to scan the channels of the polarimeter in order to calibrate all of them individually. Moreover, polarimeter modules have been brought to the European Synchrotron Radiation Facility (ESRF) in Grenoble, France, for a detailed characterization of their response to a polarized hard X-ray beam\cite{ESRF_paper}. A sub-polarimeter made of 9 polarimeter modules is foreseen to be calibrated at ESRF early 2026, while the final 100 modules will be calibrated using the setup based on scattered radioactive isotopes at CERN.

\subsection{Space Qualification}\label{subsec:space_qualif}

In order to ensure the proper functioning of the instrument once operated in space, a series of space qualification campaigns have been conducted on individual components as well as on polarimeter modules. 

\subsubsection{Irradiation}\label{subsubsec:space_qualif_irr}

\begin{figure}[h!]
\centering
 \includegraphics[height=.38\textwidth]{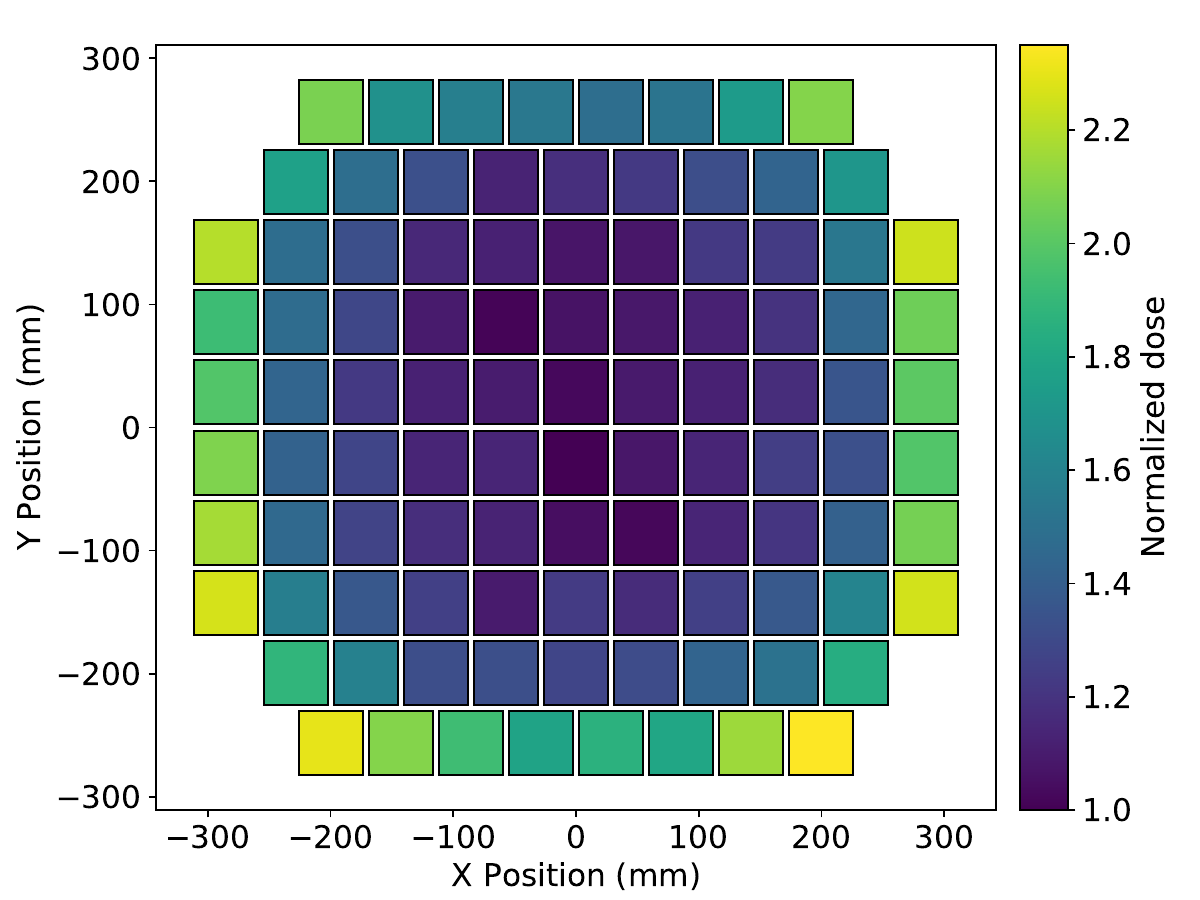}\includegraphics[height=.38\textwidth]{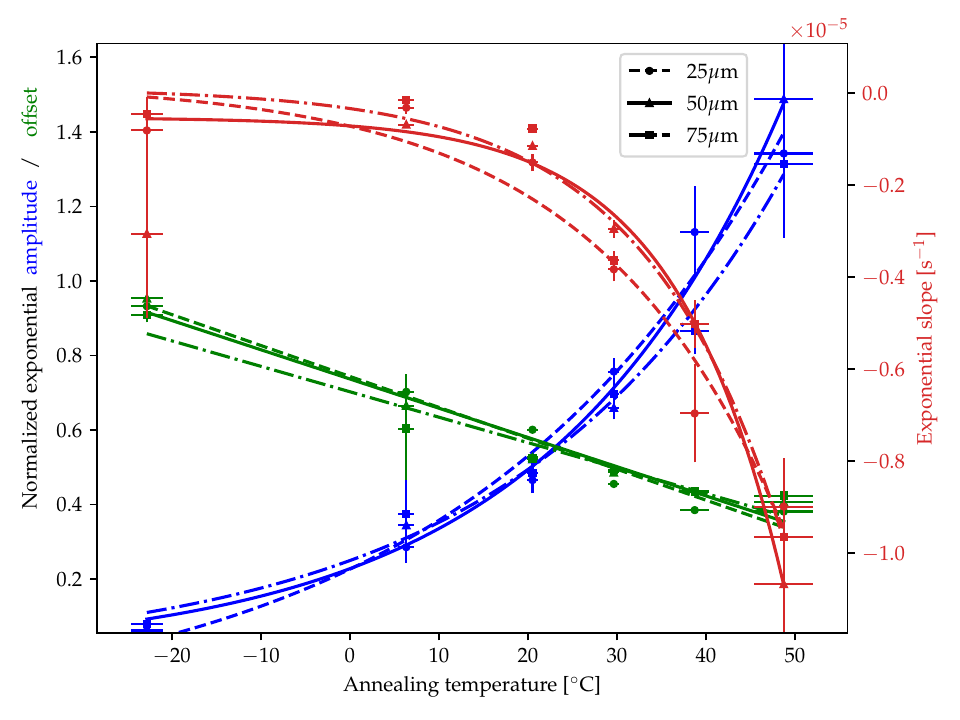}
 \caption{\textbf{Left:} Map of the expected radiation dose per polarimeter module normalized by the lowest value \citefig{SiPM_irradiation_POLAR-2}. It should be noted that the former module configuration was used here; they are now disposed in a 10$\times$10 array. \textbf{Right:} Temperature dependence of the fit parameters of the sensor dark current annealing-induced exponential recovery for Hamamatsu's S13360 SiPMs with 25, 50, and 75 $\mu$m microcell sizes \citefig{Annealing}.}
 \label{fig:SiPM_irr}
\end{figure}

Irradiation campaigns using a 58~MeV proton beam at IFJ-PAN ( Krakow, Poland) and a PuBe neutron source at NCBJ (Otwock, Poland) have been conducted to assess the radiation hardness of individual components as well as sub-system parts. The required irradiation time based on the expected dose in-orbit was computed through Geant4 simulations using well-known background models as explained in \cite{SiPM_irradiation_POLAR-2, POLAR-2_scint_paper}. A map of the 100 polarimeter modules showing the expected normalized dose in each of them is shown in Figure \ref{fig:SiPM_irr}, giving an idea of the non-uniformity of the background due to the instrument's geometry. The post-irradiation behavior of the plastic scintillators\cite{POLAR-2_scint_paper}, SiPMs\cite{SiPM_irradiation_POLAR-2, Annealing, NDA_thesis}, individual electronics components\cite{NDA_thesis, ElectronicsIrr}, and front-end electronics \cite{FEEpaper, ElectronicsIrr}, have been deeply studied to ensure that the selected items could be employed in POLAR-2's flight design.

Furthermore, thermal annealing of the radiation damage in the SiPM lattice have been studied by storing various irradiated sensors at different temperatures in order to characterize the radiation damage recovery with time as a function of temperature\cite{Annealing, NDA_thesis}. In particular, the temperature-dependent exponential decay with time of the sensor's dark current was measured, as shown in Figure \ref{fig:SiPM_irr}, which helped to define the thermal operations of the instrument. The sensors will be cooled by a Peltier element placed in each polarimeter module during normal operation to reduced the dark noise as much as possible, and will then be heated up every few months using heaters placed on the front-end electronics to perform thermal annealing of the radiation damages and partially recover their original performances.

\subsubsection{Vibration and Shock Testing}\label{subsubsec:space_qualif_vibr}

\begin{figure}[h!]
\centering
\begin{minipage}{0.35\textwidth}
    \centering
    \includegraphics[width=\textwidth]{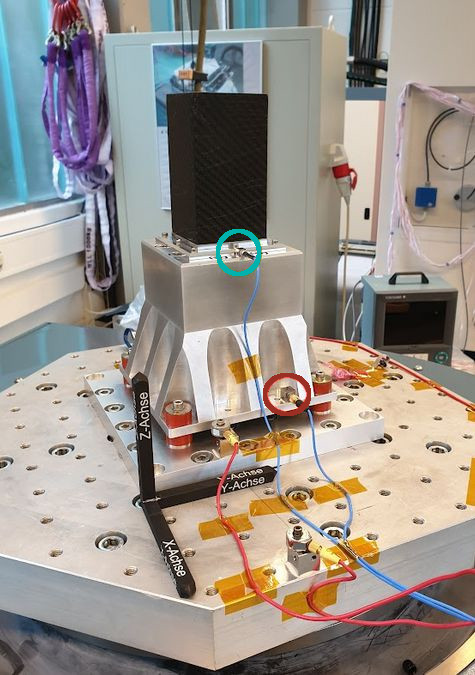}
    \vspace{-0.3cm}
    \caption{Definition of the X-Y-Z frame with respect to system. The Z axis is defined as the vertical direction (along the scintillator length), and the X and Y directions are defined in the horizontal plane along the scintillators. The readout system is placed in the aluminum frame. Accelerometers are placed on the flange of the module (\textcolor{color2}{green}) and near the dampers (\textcolor{color1}{red}) in order to monitor the resonance spectrum during and between the different tests \citefig{NDA_thesis}.}
    \label{fig:vibr_setup}
\end{minipage}\hfill
\begin{minipage}{0.62\textwidth}
    \centering
    \begin{tabular}{|c|c|c|c|c|} 
    \hline
    Frequency range [Hz] & 4-12 & 12-17 & 17-75 & 75-100 \\ \hline\hline
    Qualification requirement & 15~mm & 8.8~g & 14.5~g & 11~g \\ \hline
    \end{tabular}
    \vspace{0.1cm}   
    \caption{Requirements used for the sinusoidal vibrations qualification test performed on the 3 axis (x, y, and z), with an acceleration rate of 2~octave/minute.}
    \label{table:sinusoidal_vibration_specs}
    
    \vspace{1em}
    
    \begin{tabular}{|l|c|c|c|} 
    \hline
    & \multicolumn{3}{|c|}{Frequency range [Hz]}\\
    & 12-250 & 250-800 & 800-2000 \\ \hline\hline
    PSD & 6~dB/octave & 0.14~g$^2$/Hz & -9~dB/octave \\ \hline
    Total RMS acc. & \multicolumn{3}{|c|}{11.65~grms} \\ \hline
    Test duration & \multicolumn{3}{|c|}{180~s} \\ \hline
    Acc. directions & \multicolumn{3}{|c|}{3 axis} \\ \hline
    \end{tabular}
    \vspace{0.1cm}   
    \caption{Requirements used for the random vibrations qualification test.}
    \label{table:random_vibration_specs}
    
    \vspace{1em}
    
    \begin{tabular}{|l|c|c|} 
    \hline
    & \multicolumn{2}{|c|}{Frequency range [Hz]}\\
    & 100-500 & 500-3000 \\ \hline\hline
    Shock response spectral acceleration & 9~dB/octave & 800~g \\ \hline
    Test duration & \multicolumn{2}{|c|}{3 times per axis} \\ \hline
    Acceleration directions & \multicolumn{2}{|c|}{3 axis} \\ \hline
    \end{tabular}
    \vspace{0.1cm}   
    \caption{Requirements used for the shock qualification test.}
    \label{table:shock_specs}
    
\end{minipage}
\end{figure}

Survival to harsh launch conditions was tested at the module level through random and sinusoidal vibrations as well as shock testing along all three axes using a vibration table at MPE (Garching, Germany). The X, Y, and Z directions defined for these tests are shown in Figure \ref{fig:vibr_setup} together with the accelerometers placed on the outside of the module to monitor the system during the campaign. Tables \ref{table:sinusoidal_vibration_specs}, \ref{table:random_vibration_specs}, and \ref{table:shock_specs} respectively show the requirements imposed as an input to the sinusoidal vibration, random vibration, and shock tests.

\begin{figure}[h!]
\centering
 \includegraphics[height=.34\textwidth]{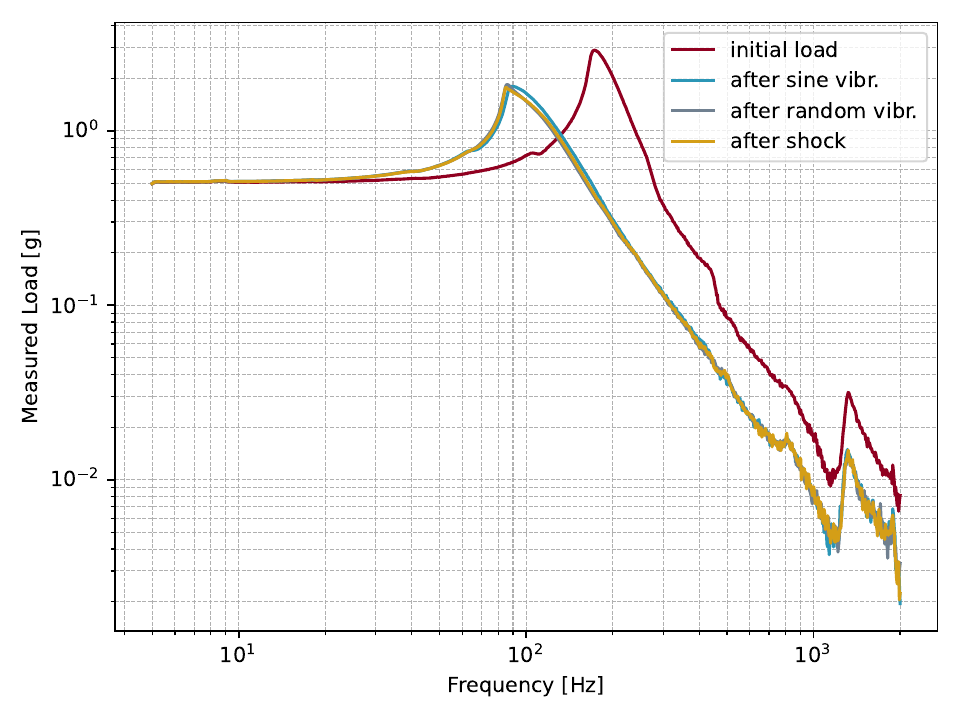}\hspace*{0.1cm}\includegraphics[height=.34\textwidth]{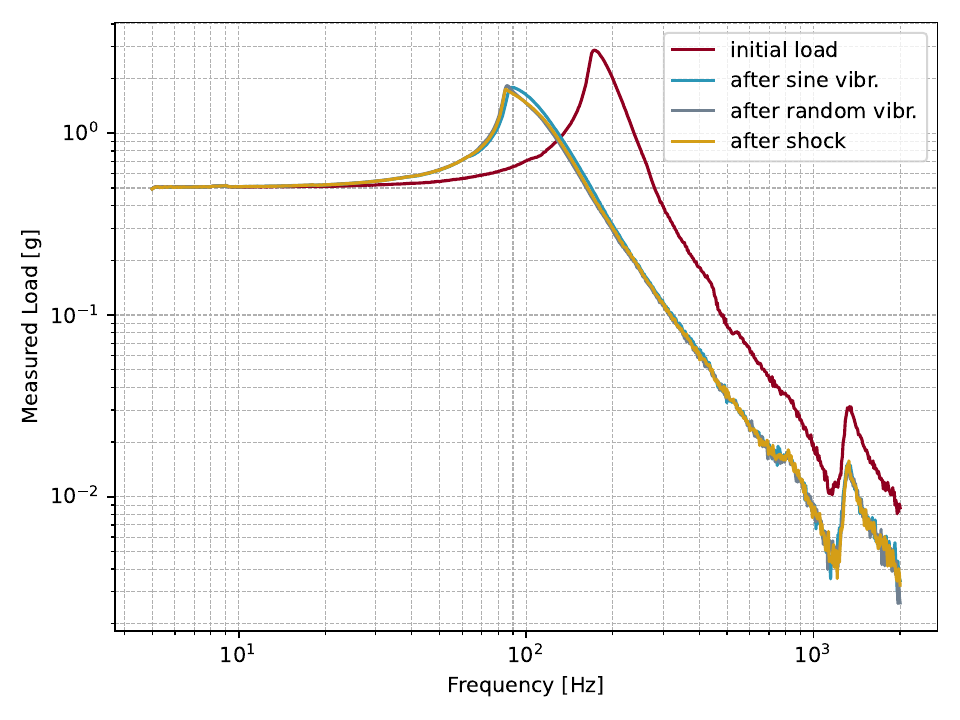}
 \includegraphics[height=.34\textwidth]{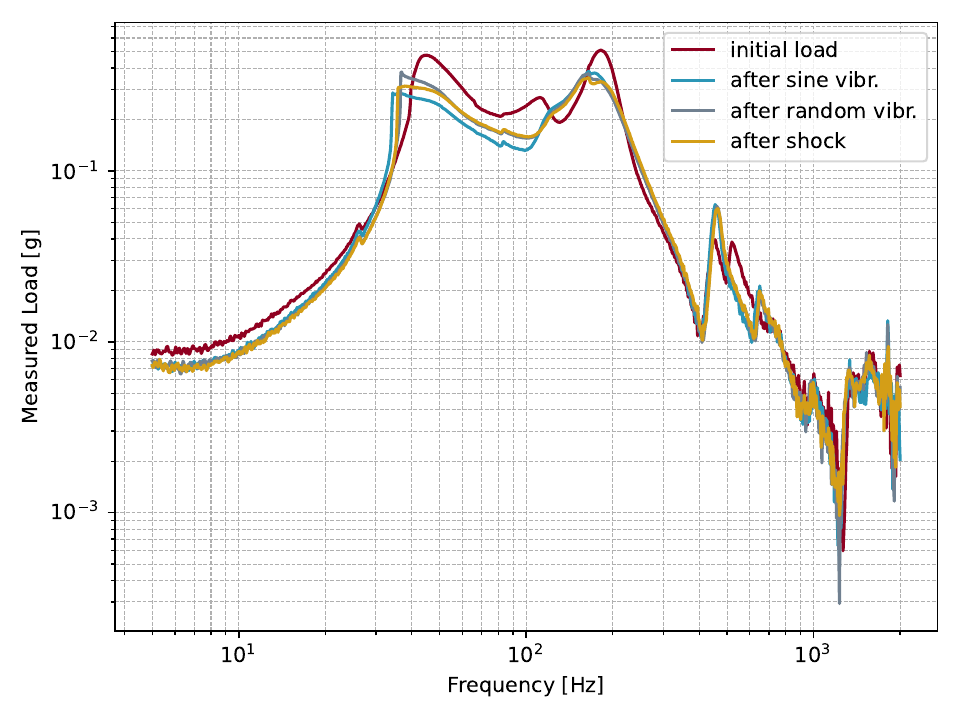}\hspace*{0.1cm}\includegraphics[height=.34\textwidth]{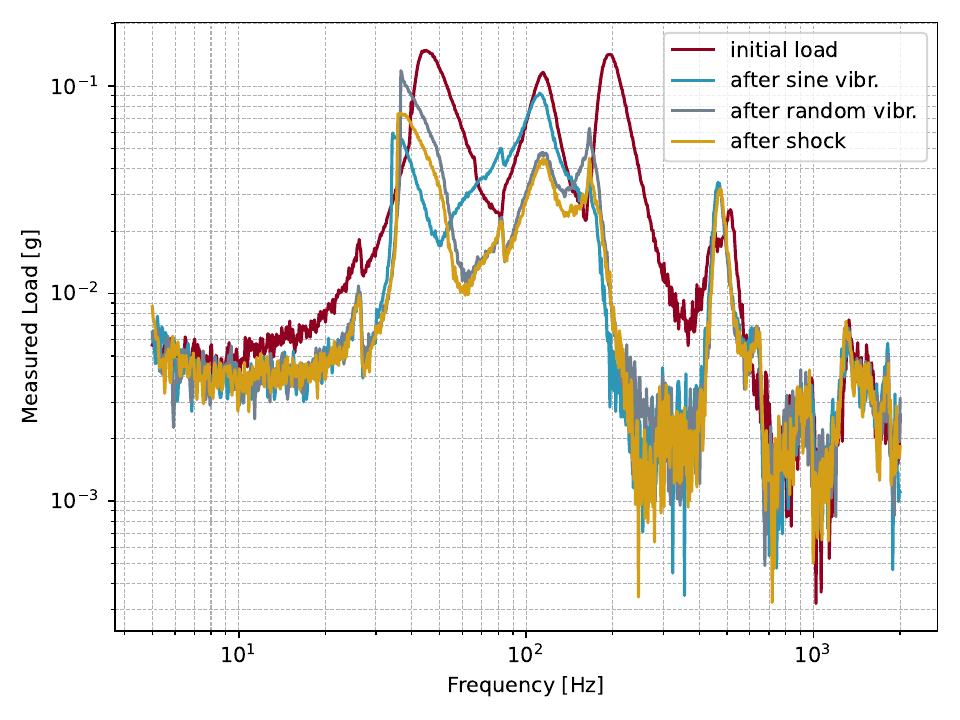}
 \includegraphics[height=.34\textwidth]{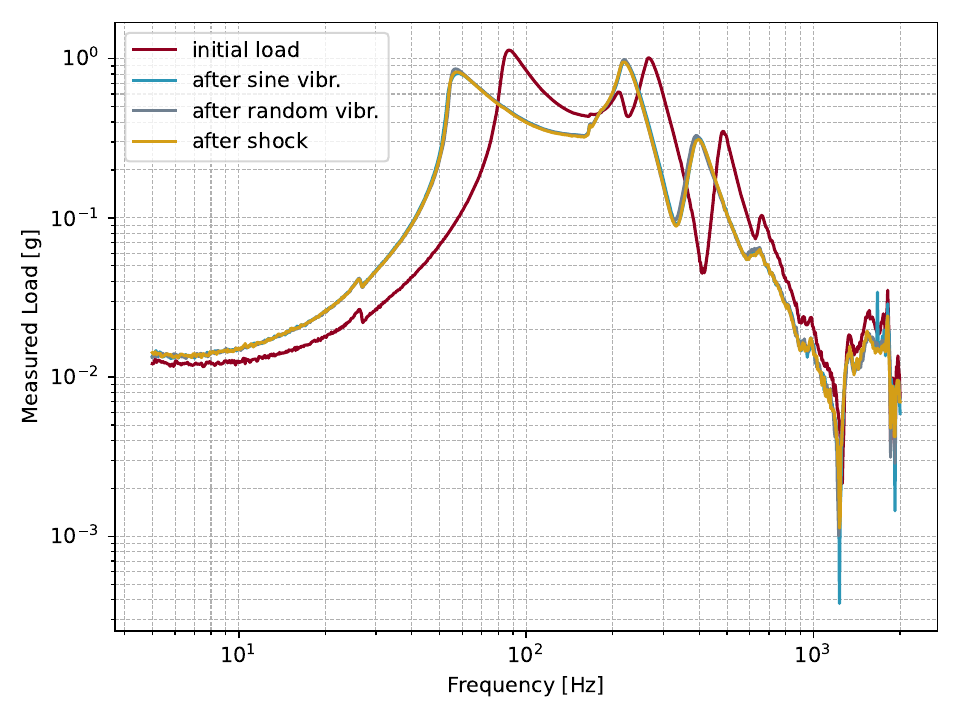}\hspace*{0.1cm}\includegraphics[height=.34\textwidth]{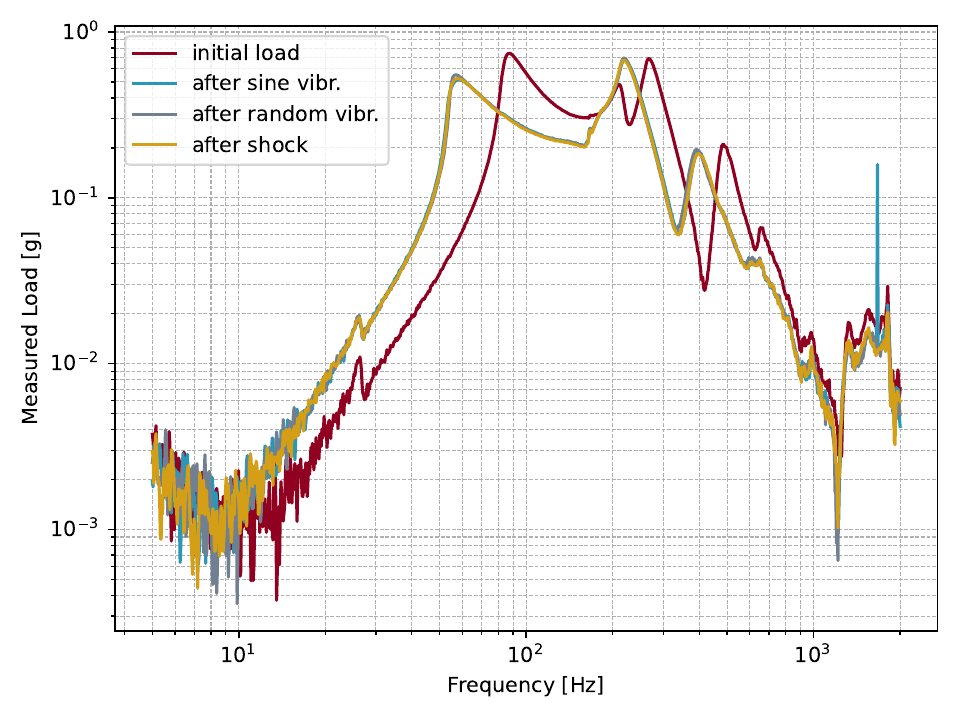}
 \caption{Load measured with the flange (\textbf{left}) and damper (\textbf{right}) accelerometers between each vibration and shock run along the z (\textbf{top}), y (\textbf{middle}), x (\textbf{bottom}) axis}
 \label{fig:vibr_load}
\end{figure}

The spectral load was measured at the start of the campaign as well as in between each test. The module was first tested on the Z-axis, then along Y, and finally along X. For each axis, the module was first vibrated with a sinusoidal, then randomly, and finally the shocks were applied. The measured spectral loads by both accelerometers along all three axis between each tests are shown in Figure \ref{fig:vibr_load}. More details about the tests and results can be found in \cite{NDA_thesis}.

\newpage

\newpage
\subsubsection{Thermal Vacuum Cycling}\label{subsubsec:space_qualif_tvt}

\begin{figure}[h!]
\centering
 \includegraphics[height=.34\textwidth]{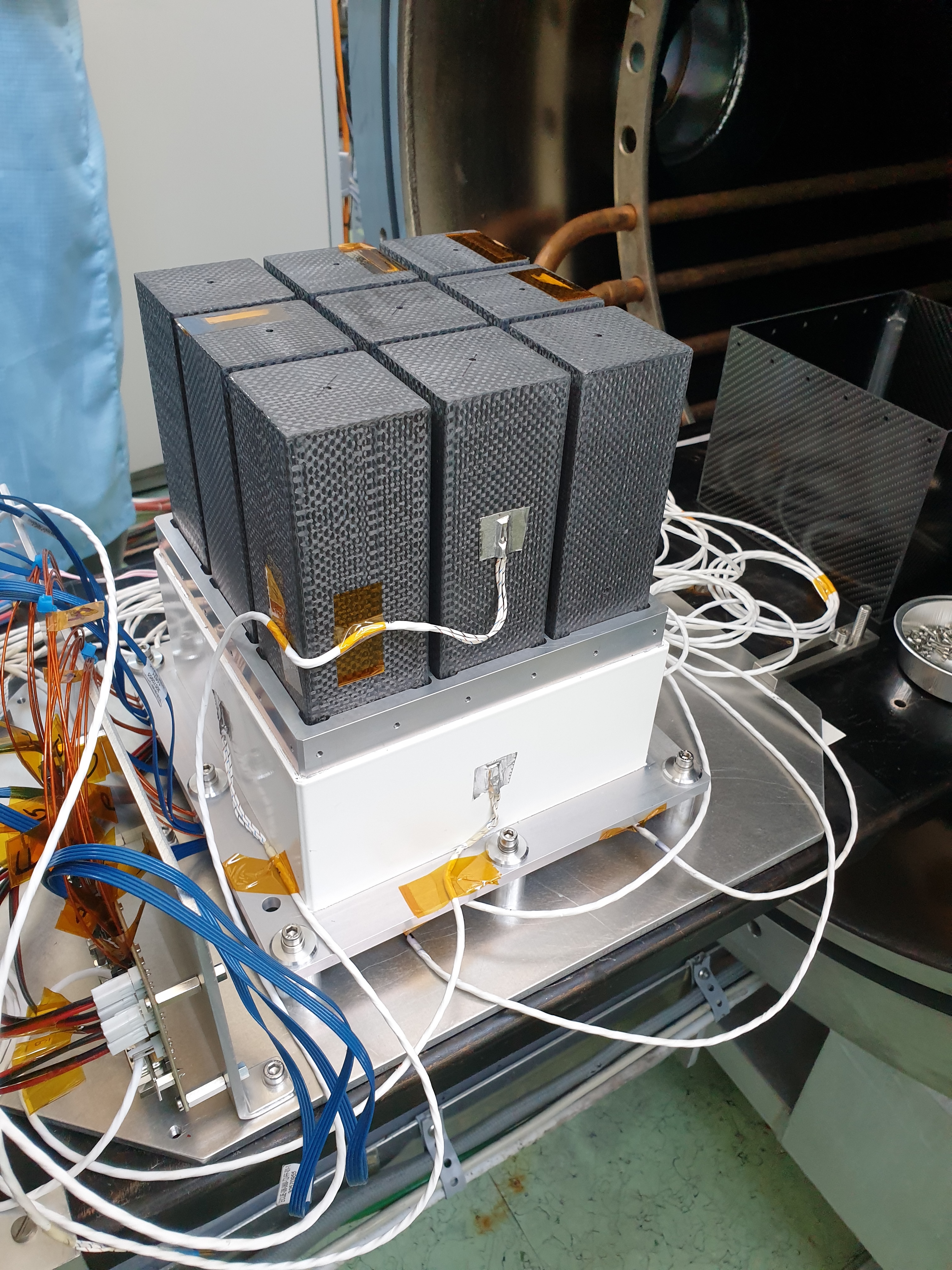}\hspace*{0.2cm}\includegraphics[height=.35\textwidth]{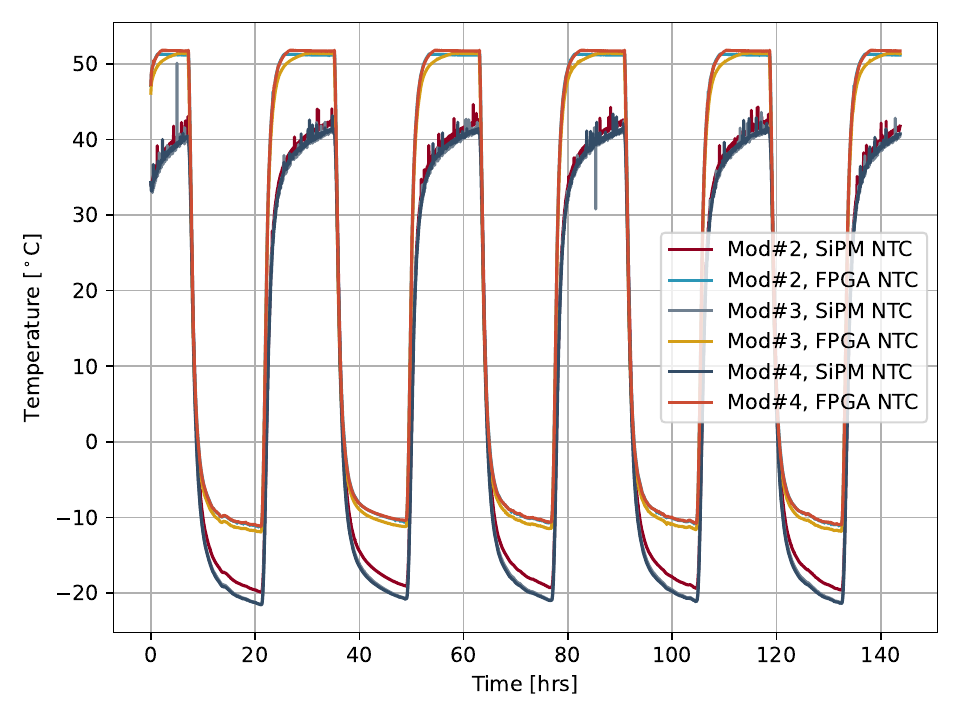}
 \caption{\textbf{Left:} Polarimeter prototype made of 4(real)+5(thermal dummy) modules being equipped with temperature sensors in the thermal chamber at MPE, Garching. \textbf{Right:} Temperature evolution of 3 real modules during thermal vacuum cycling as probed using the NTC sensors on the FEEs next to the SiPMs and FPGA.}
 \label{fig:tvt}
\end{figure}

A sub-polarimeter composed of 4 real polarimeter modules and 5 thermal dummies\footnote{Thermally and mechanically representative modules made of fake electronic boards on which thermal heaters were placed to emulated the thermal behavior of the actual flight electronics.} arranged in a 3$\times$3 array was tested for outgassing and thermally cycled in vacuum in the range of -20$^\circ$C to +50$^\circ$C for about a week \cite{NDA_thesis, FEEpaper}. The sub-polarimeter as well as the temperature monitored during several cycles for 3 of the real modules and at two locations on the electronics (near the sensors and near the FPGA, see \cite{NDA_thesis, FEEpaper} for the electronics layout) are shown in Figure \ref{fig:tvt}.

The polarimeter module design successfully passed all the irradiation, vibration and shock, and thermal vacuum qualification tests and reached a Technology Readiness Level\footnote{ECSS-E-HB-11A – Technology readiness level (TRL) guidelines -- \url{https://ecss.nl/home/ecss-e-hb-11a-technology-readiness-level-trl-guidelines-1-march-2017/} -- Consulted on 21\textsuperscript{st} July 2025.} of 7.

\newpage
\section{OUTLOOK}\label{sec:outlook}

\subsection{Expected Scientific Outcome}\label{subsec:science_outcome}

\begin{figure}[h!]
\centering
 \includegraphics[height=.42\textwidth]{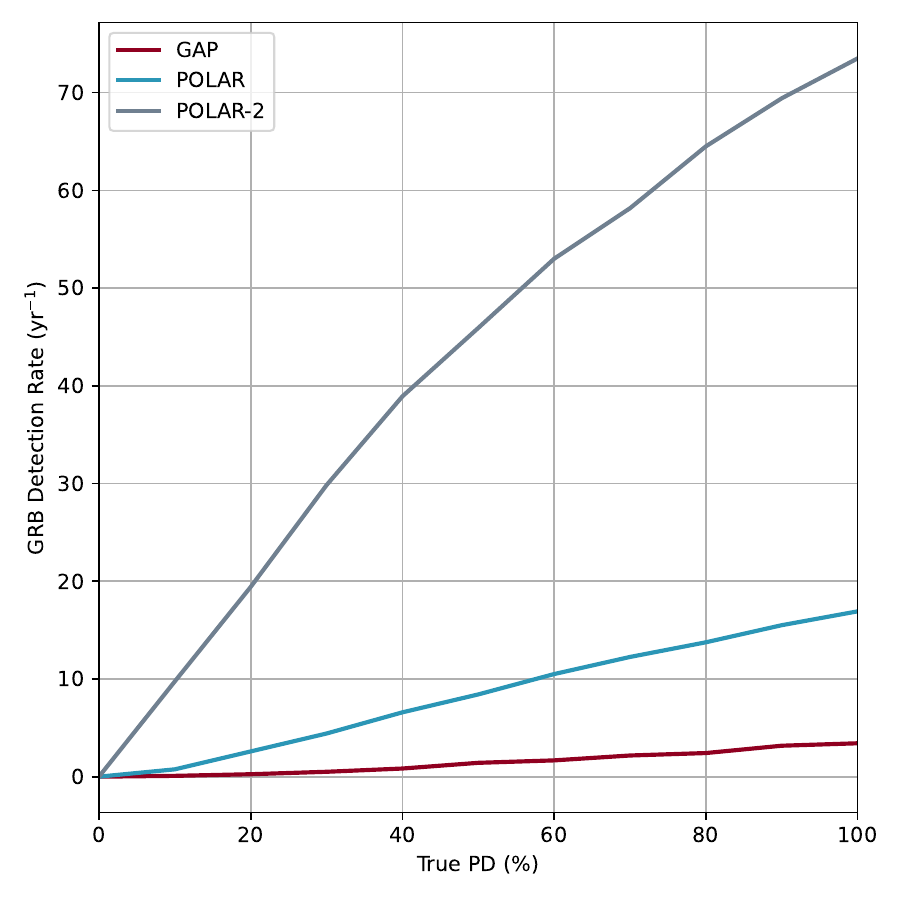} \hspace*{0.2cm}\includegraphics[height=.42\textwidth]{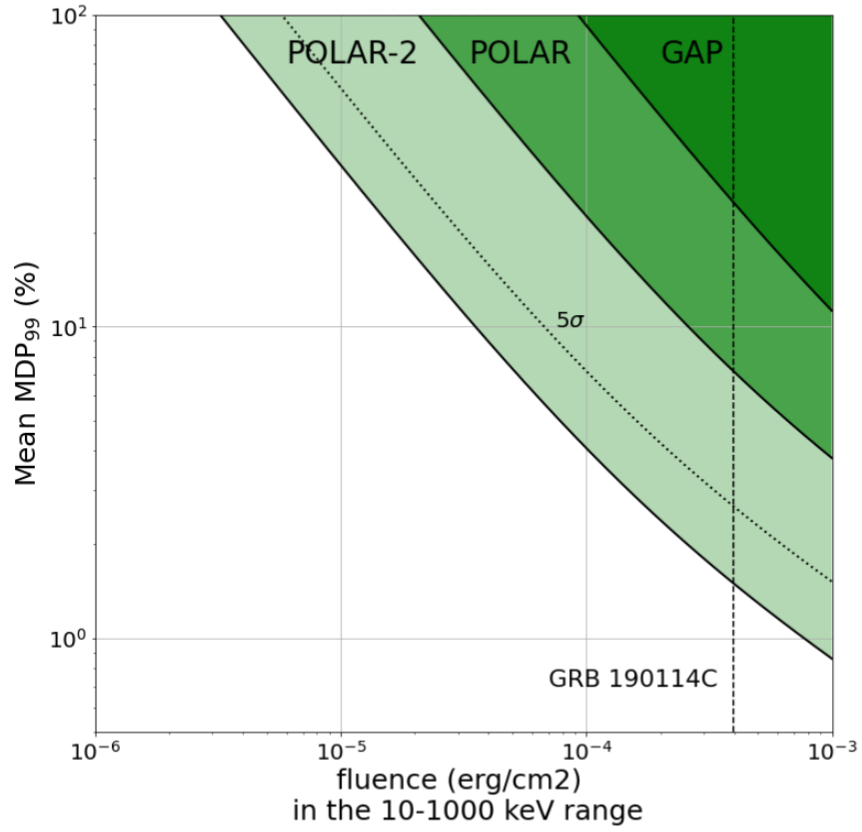}
 \caption{\textbf{Left:} GRB detection rate as a function of its true polarization for GAP, POLAR, and POLAR-2 \citefigadapt{GRBpol_review_merlin_ramandeep_joni}. \textbf{Right:} Minimum detectable polarization with 99~\% confidence level for a 100~s long GRB as a function of the burst fluence in the 10-1000~keV band for GAP, POLAR, and POLAR-2 \citefig{GRBpol_review_merlin_ramandeep_joni}.}
 \label{fig:science_perf}
\end{figure}

POLAR-2 is expected to detect about 50 GRBs per year with at least the same statistics as for the brightest bursts observed by POLAR, enabling the possibility of performing temporal and energy-resolved polarimetric analysis for a considerable population of bursts. The expected yearly rate of bursts for POLAR-2 as a function of its true polarization is compared to the rates from GAP and POLAR in Figure \ref{fig:science_perf}. This Figure also shows the Minimum Detectable Polarization at a 99~\% confidence level\cite{weisskopf_x-ray_polarization, Strohmayer_2013} for a long GRB as a function of the burst's fluence for POLAR-2, compared with GAP and POLAR. A significant improvement in sensitivity can be observed.

Convolving the effective area of POLAR-2 for polarimetry with a typical GRB spectrum, the amount of photons detected for a given burst was improved by one order of magnitude compared to POLAR, despite increasing the number of module by only a factor 4. This was made possible by the technological upgrades and design optimization brought to the POLAR-2 polarimeter module described in Section \ref{sec:polar-2}. We aim at implementing a similar rapid alert system via the Beidou satellite constellation as implemented for Einstein Probe. This would allow POLAR-2 to significantly contribute to the field of multi-wavelength astronomy by enabling the afterglow follow-up of GRBs thanks to a fast alerting system sending to ground within a minute preliminary localization of the burst. Deep learning algorithms have been studied as a way to perform fast on-board pre-localization of the bursts \cite{HAGRIDproceeding, Gilles_HAGRID_master}. Moreover, as POLAR-2 will be operated during LIGO's O5 run, it could also contribute to multi-messenger astronomy. A joint Gravitational Wave detection has the potential to rule out most of the existing models by observing a single burst by combining the merger parameters obtained from GW measurements with the polarimetric parameters measured by POLAR-2 \cite{Kole_MM_GRB}.

\subsection{Current Status of POLAR-2}\label{subsec:status}

POLAR-2 is approved for a launch to the China Space Station (CSS) and has entered its construction phase. After several years of design, development, and testing of the polarimeter modules, the construction of the flight polarimeter has started. A first sub-polarimeter will be tested at ESRF early Februeary 2026, while the full instrument should be delivered for acceptance tests early 2027 for a launch mid/end 2027 to the CSS. In addition to this Compton polarimeter, a Broad-band Spectrometer (BSD) developed by IHEP (Beijing, China) is under consideration to enhance POLAR-2's spectral capabilities \cite{BSD}. A third type of instrument based on gas photoelectric polarimeters, the Low-energy Polarization Detector (LPD), is being developed by the University of GuangXi (Nanning, China) to extend the polarimetric capabilities of POLAR-2 down to the X-ray band\cite{LPD1, LPD2}.

\acknowledgments 

We gratefully acknowledge the Swiss Space Office of the State Secretariat for Education, Research and Innovation (ESA PRODEX Programme) which supported the development and production of the POLAR-2 detector. M.K and N.D.A acknowledge the support of the Swiss National Science Foundation. National Centre for Nuclear Research acknowledges support from Polish National Science Center under the grant UMO-2018/30/M/ST9/00757. We gratefully acknowledge the support from the National Natural Science Foundation of China (Grant No. 11961141013, 11503028), the Xie Jialin Foundation of the Institute of High Energy Phsyics, Chinese Academy of Sciences (Grant No. 2019IHEPZZBS111), the Joint Research Fund in Astronomy under the cooperative agreement between the National Natural Science Foundation of China and the Chinese Academy of Sciences (Grant No. U1631242), the National Basic Research Program (973 Program) of China (Grant No. 2014CB845800), the Strategic Priority Research Program of the Chinese Academy of Sciences (Grant No. XDB23040400), and the Youth Innovation Promotion Association of Chinese Academy of Sciences.

\bibliography{BibliographyPOLAR} 
\bibliographystyle{spiebib} 

\end{document}